\documentstyle[sprocl]{article}

\input{psfig.sty}

\def\Journal#1#2#3#4{{#1} {\bf #2}, #3 (#4)}

\def\NPB{{\em Nucl. Phys.} B}
\def\PLB{{\em Phys. Lett.}  B}

\def\PRD{{\em Phys. Rev.} D}

\def\be{\begin{equation}}
\def\ee{\end{equation}}
\def\bea{\begin{eqnarray}}
\def\eea{\end{eqnarray}}
\def\slash#1{\rlap{$#1$}/} % slashes a character
\def\dsl{\,\raise.15ex\hbox{/}\mkern-13.5mu D} %this one can be subscripted
\def\delsl{\raise.15ex\hbox{/}\kern-.57em\partial}
\def\grad#1{\,\nabla\!^{{#1}}\,}

\def\Tr{{\rm Tr}}
\def\N{N_c}
\def\L{{\cal L}}

\def\S{{\cal S}}

\def\D{{\cal D}}
\def\A{{\cal A}}
\def\M{{\cal M}}

\def\H{{\cal H}}

\def\V{{\cal V}}

\def\ltap{\ \raise.3ex\hbox{$<$\kern-.75em\lower1ex\hbox{$\sim$}}\ }
\def\gtap{\ \raise.3ex\hbox{$>$\kern-.75em\lower1ex\hbox{$\sim$}}\ }
\def\ssqr#1#2{{\vbox{\hrule height #2pt
      \hbox{\vrule width #2pt height#1pt \kern#1pt\vrule width #2pt}
      \hrule height #2pt}\kern- #2pt}}

\def\bsqr{\ssqr{10}{.1}}
\def\nbox{\hbox{$\bsqr\bsqr\bsqr\bsqr\raise2.7pt\hbox{$\,\cdot\cdot
\cdot\cdot\cdot\,$}\bsqr\bsqr\bsqr$}}

\def\openone{{\bf 1}}
\begin{document}

\title{BARYON CHIRAL PERTURBATION THEORY IN THE 1/$\bf \N$ 
EXPANSION}

\author{E. JENKINS}

\address{Department of Physics, 9500 Gilman Drive, University of California at
San Diego, 
La Jolla,\\ CA 92093-0319, USA\\E-mail: ejenkins@ucsd.edu} 


\maketitle\abstracts{ The chiral Lagrangian for baryons is formulated in 
an expansion in $1/N_c$.  The chiral Lagrangian implements the contracted 
spin-flavor symmetry of large-$N_c$ baryons as well as nonet symmetry of 
the leading planar diagrams.  Large-$N_c$ consistency conditions ensure 
that chiral loop corrections are suppressed in $1/N_c$ through exact 
cancellation of chiral loop graphs to fixed orders in $1/N_c$.  Application 
of $1/N_c$ baryon chiral perturbation theory to the flavor-$\bf 27$ baryon 
mass splittings and the baryon axial vector currents are considered as 
examples.}

\section{Introduction}

The realization that large-$N_c$ baryons respect a contracted spin-flavor
symmetry has resulted in many important predictions for the spin and
flavor properties of baryon amplitudes~\cite{dm,j,annrev,zacatecas}.  
The $1/N_c$ corrections to the
large-$N_c$ limit have been classified for general spin and flavor
representations.  
Characterization of the $1/N_c$ breakings of large-$N_c$ baryon
spin-flavor symmetry has been essential for the application of the $1/N_c$
expansion to the case of physical interest, QCD baryons with $N_c =3$.  

The study of QCD baryons in an expansion in $1/N_c = 1/3$ has been quite
successful phenomenologically.  $1/N_c$
breaking of contracted $SU(6)$ spin-flavor symmetry is comparable in magnitude
to the breaking of $SU(3)$ flavor symmetry.
This fact makes it imperative to 
reformulate baryon chiral perturbation theory in an expansion in $1/N_c$.

In this talk, I will describe the progress which has been made in 
baryon chiral perturbation theory using the $1/N_c$ expansion.  The 
chiral Lagragian for baryons is formulated in an expansion in $1/N_c$ and
$m_q/ \Lambda_{\rm QCD}$ about the double limit 
$N_c \rightarrow \infty$, $m_q/ \Lambda_{\rm QCD} \rightarrow 0$, with
the ratio ${1 \over {N_c}}/ \left(m_q/ \Lambda_{\rm QCD} \right)$ held
fixed~\cite{1/nbcl}.
In this double expansion, the chiral limit cannot be taken independently of the 
$1/N_c \rightarrow 0$ limit, so there is no order-of-limits ambiguity 
associated with the $1/N_c$ baryon chiral Lagrangian.

\section{Heavy Baryon Chiral Perturbation Theory}
For large, finite $N_c$, a baryon has mass $M$ which is
${\cal O}(N_c)$ while 
mesons have masses which are ${\cal O}(1)$.  Thus, a large-$N_c$
baryon acts as a heavy 
static fermion in its interactions with pions.  The momentum of an on-shell 
baryon which absorbs a pion carrying momentum $k$ can be written
as
\begin{equation}\label{mv}
P^\mu = M v^{\prime \mu} = M v^\mu + k^\mu \ ,
\end{equation}
where $v^\prime$ and $v$ are the velocities of the initial and 
final baryons.  It readily follows that
the baryon velocity is conserved in its interactions with
pions,
\begin{equation}
v^{\prime \mu} = v^\mu + O(1/N_c)\ .
\end{equation}
Thus, it is convenient to formulate
Heavy Baryon Chiral Perturbation Theory (HB$\chi$PT)~\cite{hbcpt,jmhungary}
in terms of the velocity-dependent baryon field
\begin{equation}\label{bv}
B_v(x) = e^{iM \slash v \ v_\mu x^\mu}\left({{1 + \slash v} \over 2}\right) 
B(x)\ . 
\end{equation}
In the rest frame of the baryon, Eq.~(\ref{bv})
redefines the baryon field to include the phase factor $e^{iMt}$ and projects
onto the particle portion of the Dirac spinor.  The baryon propagator in HB$\chi$PT is
given by 
\begin{equation}
{{i \left( \slash P + M \right)} \over {P^2 - M^2}} \rightarrow
{i \over {k \cdot v} } \left( {{1 + \slash v} \over 2} \right)\ ,
\end{equation}  
which reduces to $i/E$ in the rest frame of the baryon.

A number of interesting phenomenological results were found
in HB$\chi$PT: (1) Chiral loop corrections are large if only spin-$\frac 1 2$
octet baryons are included in the heavy baryon chiral Lagrangian.  
(2) Including spin-$\frac 3 2$ decuplet baryons results in significant cancellation 
between loop graphs with intermediate octet and decuplet states.  
(3) The phenomenological fit to data yields $SU(6)$-like
couplings for the parameters of the heavy
baryon chiral Lagrangian.  All of these 
features are not consequences of the $SU(3)$ flavor symmetry
of the baryon chiral Lagrangian, but are explained by large-$N_c$ spin-flavor
symmetry.

In the large-$N_c$ limit, the baryon spin-flavor symmetry
requires inclusion of the complete spin-flavor multiplet ${\bf 56}$ which
contains both the spin-$\frac 1 2$ octet and spin-$\frac 3 2$ decuplet baryons.  
The decuplet-octet mass splitting
\begin{equation}
\Delta \equiv \left( m_T -
m_B \right) \sim 1/N_c 
\end{equation}
vanishes in the large-$N_c$ limit, so
the spin-$\frac 1 2$ and spin-$\frac 3 2$ baryons
are degenerate.  Including only the spin-$\frac 1 2$ octet
baryons in the chiral Lagrangian breaks large-$N_c$ baryon spin-flavor symmetry
explicitly, causing chiral loops to {\it grow} with powers of $N_c$.  In
contrast, chiral loops for the ${\bf 56}$ are suppressed by powers of $1/N_c$.  
The proper $1/N_c$ power counting of loop corrections is restored due to
exact large-$N_c$ cancellations among different loop diagrams.
In addition, the $1/N_c$ expansion 
provides a {\it quantitative} understanding of spin-flavor
symmetry for baryon couplings.

\section{Chiral Loop Cancellations}
\begin{figure}
\psfig{figure=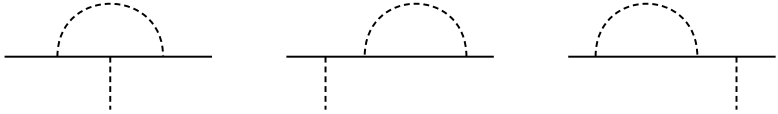,height=1.5in}
\vskip-0.85truein
\caption{Diagrams which contribute to the baryon axial-vector pion coupling 
at one loop.  Individual diagrams are order $N_c$ times the tree-level coupling,
whereas the sum of all the diagrams is order $1/N_c$ times the tree-level 
coupling.}
\end{figure} 

The diagrams which correct the baryon axial vector pion-coupling at one loop
are displayed in Fig.~1.  Each baryon--pion vertex
is order $\sqrt{N_c}$, so each loop graph with three vertices is order
$N_c^{3/2}$, i.e. a factor of $N_c$ larger than the
tree-level vertex.  It is possible to show, however, that
the sum of the graphs is proportional to a double
commutator of three baryon axial vector currents,
\begin{equation}
\left( \sqrt{N_c} \right)^3 \left[ X^{jb}, \left[ X^{jb}, X^{ia} \right]
\right] \le {\cal O} \left( { 1 \over {\sqrt{{N_c}}} } \right),
\end{equation}
which is at least a factor of
$1/N_c$ smaller than the tree-level vertex. 
The large-$N_c$ consistency condition 
appearing in the one-loop chiral correction is precisely
the same consistency condition obtained from the analysis of tree diagrams for
$B + \pi \rightarrow B^\prime + \pi + \pi$
scattering.

In general, exact large-$N_c$ 
cancellations ensure that the chiral
correction at $L$ loops is of relative order $(1/N_c)^L$, 
rather than order $(N_c)^L$.
Notice that the cancellations are
increasingly significant as the number of loops $L$ increases.  

\section{``Nonet'' Symmetry}
For finite and large $N_c$, planar diagrams dominate the quark-gluon
dynamics~\cite{thooft}.  
Quark loops are suppressed by one power of $1/N_c$,
so there is no quark-antiquark pair creation and annihilation at leading
order.  The suppression of quark loops in large-$N_c$ QCD implies that planar
QCD has the additional flavor symmetry~\cite{veneziano}
\begin{equation}
U(1)_{q_i} \times U(1)_{\bar q_i}\ ,
\end{equation}
which says that the number of quarks and the number of antiquarks
of each flavor are separately conserved by the leading planar diagrams.  This
planar flavor symmetry implies that there are nonets of mesons in large $N_c$, i.e.
the $\pi$, $K$, $\eta$ and $\eta^\prime$ form a flavor nonet in the large-$N_c$
limit, as well as Zweig's rule.  The consequences of planar flavor symmetry
for baryons, however, have been obtained only recently~\cite{1/nbcl}.

In the large-$N_c$ planar limit, baryon diagrams consist of $N_c$ quark lines
connected by planar gluon exchange with no quark loops.  The $SU(3)$ flavor
symmetry of QCD extends to a $U(3)$ flavor symmetry for large-$N_c$ QCD in the
planar limit.  For baryons, $U(3)$ planar flavor symmetry implies that baryon 
amplitudes form
representations of $U(3)$ flavor symmetry up to a correction of relative
order $1/N_c$.
For example, the baryon axial vector flavor-octet currents $A^{ia}$
and the baryon axial vector flavor-singlet current $A^i$ form a flavor nonet 
at leading order,
\begin{equation}
A^i = A^{i9} + {\cal O} \left( {1 \over N_c} \right) \ .
\end{equation}
This nonet symmetry is broken at relative order $1/N_c$ by
nonplanar baryon diagrams with a single quark loop, Fig.~2.
Note that $U(3)$ flavor symmetry is a symmetry of large-$N_c$ QCD in the leading
planar limit only; it is not a symmetry of QCD itself.  

\begin{figure}
\psfig{figure=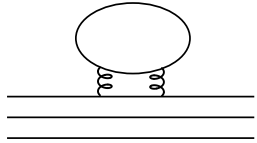,height=1.0in}
%{\includegraphics{b_fig5.ps}}
\caption{Baryon diagram with a single quark loop.  The diagram is suppressed by
relative order $1/N_c$ compared to the leading planar diagrams with no quark
loops.  Note that quark number and antiquark number are not separately conserved
by the diagram.}
\end{figure} 

\section{$1/N_c$ Baryon Chiral Lagrangian}

It is natural to formulate the $1/N_c$ baryon chiral Lagrangian 
in the rest frame of the baryon.  In compact notation, the
chiral Lagrangian is given in the flavor symmetry limit by
\begin{equation}\label{bary}
\L_{\rm baryon} = i \D^0 - M_{\rm hyperfine}
+ \Tr \left( \A^i \lambda^a \right) A^{ia}
+ \Tr \left( \A^i {{2I} \over \sqrt{6}} \right) A^{i} +
\ldots, 
\end{equation}
where each of the operators is understood to act on the baryon ${\bf 56}$, and
the ${\cal O}(N_c)$ flavor-singlet mass of the baryon ${\bf 56}$ has been removed
from the chiral Lagrangian by the heavy-baryon field redefinition.  The
Lagrangian retains the hyperfine mass splitting
\begin{equation}
M_{\rm hyperfine} = \Delta {1 \over \N} J^2 \ ,
\end{equation} 
where the coefficient $\Delta$ is the mass splitting of the spin-$\frac 3 2$ and
spin-$\frac 1 2$ baryons.  The chiral Lagrangian depends on the pion fields through
the combinations
\begin{equation}\label{picurrents}
\V^0 = {1 \over 2} \left( \xi \partial^0 \xi^\dagger
+ \xi^\dagger \partial^0 \xi \right),\quad
\A^i = {i \over 2} \left( \xi \grad i \xi^\dagger
- \xi^\dagger \grad i\xi \right)= \grad i \Pi / f + \cdots,
\end{equation}
which are defined in terms of 
$\xi =e^{i \Pi / f}$, where
$\Pi= {{\pi^a \lambda^a} \over 2} + {\eta^\prime \over \sqrt{6}}$ contains
the $\pi$, $K$, $\eta$ and $\eta^\prime$ fields.  Eq.~(\ref{bary})
states that the flavor-octet axial vector
pion current is coupled to the baryon axial vector flavor-octet
current $A^{ia}$, whereas
the flavor-singlet axial vector $\eta^\prime$ current is coupled to the baryon
axial vector flavor-singlet current $A^i$.  $U(3)$ planar flavor symmetry
implies that these baryon axial vector currents form a nonet at leading order in
the $1/N_c$ expansion and that 
this flavor-nonet baryon axial vector current couples to
the axial vector flavor-nonet current of pseudo-Goldstone bosons involving the
pion octet and the $\eta^\prime$ singlet.
 
Each of the baryon operators in the chiral Lagrangian
has a $1/N_c$ expansion in terms of operator
products of the baryon spin-flavor generators
\begin{equation}  
J^i = q^\dagger \left({ \sigma^i \over 2} \otimes 
\openone \right) q , \quad
T^a = q^\dagger \left(\openone \otimes {\lambda^a \over 2} \right) q, \quad
G^{ia} = q^\dagger \left({\sigma^i \over 2}
\otimes {\lambda^a \over 2}\right) q \ .
\end{equation}
It will be easy to impose the 
$U(3)$ planar flavor symmetry of baryon amplitudes
using the additional generators
\begin{equation}
G^{i9} = q^\dagger \left( {\sigma^i \over 2} \otimes {I \over \sqrt{6}} \right)
q = {1 \over \sqrt{6}} J^i , \quad
T^9 = q^\dagger \left( I \otimes {I \over \sqrt{6}} \right)q =
{1 \over \sqrt{6}} N_c \openone , 
\end{equation}
which reduce to the baryon spin generator and identity operator, up to
normalization factors.

The $1/N_c$ expansion of the axial vector flavor-octet current for QCD baryons
is given
by 
\begin{eqnarray}\label{aia}
A^{ia} = &&a_1 G^{ia} + b_2 {1 \over N_c} J^i T^a 
+ b_3 {1 \over N_c^2} \left\{ J^i, \left\{ J^j, G^{ja} \right\} 
\right\} \\
&&+ d_3 {1 \over N_c^2} \left( \left\{ J^2, G^{ia} \right\} - \frac 1 2
\left\{J^i,\left\{ J^j, G^{ja} \right\} \right\} \right)\ , \nonumber
\end{eqnarray}
where the $1/N_c$ expansion extends only to $3$-body operators for QCD baryons 
with three valence quarks.
The $1/N_c$ expansion for $A^{ia}$ is given
in terms of four unknown coefficients $a_1$, $b_2$, $b_3$ and $d_3$.  
The baryon axial vector flavor-singlet current has a $1/N_c$ expansion given by
\begin{equation} 
A^{i} = c_1 J^i + c_2 {1 \over N_c^2} \left\{ J^2, J^i \right\}\ ,
\end{equation}
in terms of two $1/N_c$ operators.  
In contrast, the parametrization of baryon axial vector flavor-octet
current in terms of
the $SU(3)$ flavor invariants of the HB$\chi$PT
Lagrangian is given by
\begin{eqnarray}
&&2 D\ \Tr\ \bar B S^\mu \left\{ {\cal A}_\mu, B \right\}
+2 F\ \Tr\ \bar B S^\mu \left[ {\cal A}_\mu, B \right] \\
&&+ C\ \left( \bar T^\mu {\cal A}_\mu B + \bar B {\cal A}_\mu T^\mu 
\right)
+ 2 H\ \bar T^\mu S^\nu {\cal A}_\nu T_\mu \ , \nonumber
\end{eqnarray}
in terms of the four $SU(3)$ parameters $D$, $F$, $C$ and $H$,
whereas the $SU(3)$ analysis of the flavor-singlet
yields two invariants
\begin{equation}
2 S_B\, \Tr \A_\mu \, \Tr \overline B_v \S_v^\mu B_v 
- 2 S_T\, \Tr \A_\nu \, \overline T_v^\mu \S_{v}^{\nu} T_{v\,\mu}\ .
\end{equation}
Notice that
there are the
same number of coefficients in the $1/N_c$ expansion as there are parameters in
the $SU(3)$ analysis; however, the $1/N_c$ expansion is advantageous because
it orders the terms in powers of $1/N_c$.  
%The parameters $D$, $F$, $C$ 
%and $H$ are related to the $1/N_c$ coefficients through   
%\begin{eqnarray}\label{dfch}
%&&D = \frac 1 2 a_1 + \frac 1 6 b_3, \nonumber\\
%&&F = \frac 1 3 a_1 + \frac 1 6 b_2
%+ \frac 1 9 b_3, \nonumber\\
%&&C = -a_1 - \frac 1 2 d_3, \\
%&&H = -\frac 3 2 a_1 - \frac 3 2 b_2
%-\frac 5 2 b_3 . \nonumber
%\end{eqnarray}
%The $SU(3)$ parameters $S_B$ and $S_T$ are related to the $1/N_c$ coefficients
%through
%\begin{equation}
%S_B = {1 \over \sqrt{6}} \left( c_1 
%+ {1 \over 6} c_3
%\right), \qquad
%S_T = {3 \over \sqrt{6}} \left( c_1 
%+ {5 \over 6} c_3 \right).
%\end{equation}

At leading order in the $1/N_c$ expansion, the following truncations are
possible, yielding relations between the $SU(3)$ parameters at leading order
in the $1/N_c$ expansion:
\begin{eqnarray} 
M = m_0 N_c \openone \quad &\Rightarrow& \quad
m_T = m_B \nonumber\\
A^{ia} = a_1 G^{ia} + b_2 {1 \over N_c} J^i T^a
\quad &\Rightarrow& \quad C = -2 D , \  H=3D-9F \\
A^i = c_1 J^i \quad &\Rightarrow& \quad 
S_B = \frac 1 3 S_T \ . \nonumber
\end{eqnarray}

In the planar limit with zero quark loops, 
$U(3)$ planar flavor implies that the baryon axial vector
flavor-singlet current $A^i$ is the ninth component of the flavor-octet currents
$A^{ia}$ so
\begin{equation}\label{axialnine} 
A^i = {1 \over \sqrt{6}}\left( a_1 + b_2 \right) J^i 
+ {1 \over \sqrt{6}}\left( 2 b_3 \right)
{1 \over N_c^2} \left\{ J^2, J^i \right\}\ ,
\end{equation}
up to a correction of relative order $1/N_c$.
Thus, $U(3)$ planar flavor
symmetry relates the {\it coefficients} of the flavor-singlet current to the
coefficients of
the corresponding flavor-octet currents at leading order in $1/N_c$.  
Note that nonet symmetry is
valid for all of the operators in the $1/N_c$ expansion separately, 
not just the leading ones,
since violation of the nonet symmetry only comes from diagrams with 
quark loops.  In terms
of the $SU(3)$ parameters, nonet symmetry implies that
\begin{equation}
S_B \rightarrow \frac 1 3 \left( 3F - D \right), 
\ S_T \rightarrow - \frac 1 3 H \ .
\end{equation}

\subsection{$SU(3)$ Flavor Symmetry Breaking}

A similar analysis can be performed for the flavor symmetry breaking terms of the
$1/N_c$ baryon chiral Lagrangian.

The $1/N_c$ baryon chiral Lagrangian with explicit $SU(3)$ flavor symmetry
breaking through the quark mass matrix is given to linear order in the quark
mass matrix ${\cal M}$ by
\begin{eqnarray}
\L^{\M}_{\rm baryon} &&=
\Tr \left( \left( \M \overline \Sigma
+ \M^\dagger \overline \Sigma^\dagger \right)
{I \over \sqrt{6}} \right) \H^0 \\ 
&& +\sum_{a=3,8} \Tr \left( \left( \overline \xi \M \overline \xi 
+ \overline \xi^\dagger \M^\dagger \overline \xi^\dagger \right)
{\lambda^a \over 2} \right)
\H^a \ ,\nonumber  
\end{eqnarray}
where the baryon flavor-singlet and flavor-octet operators ${\cal H}^0$ and
${\cal H}^a$ have $1/N_c$ expansions
\begin{eqnarray}
\H^0 &&= c_0 \N \openone + c_2 {1 \over \N} J^2  \nonumber\\
\H^a &&= d_1 T^a + d_2 {1 \over \N} \{ J^i, G^{ia} \} +
d_3 {1 \over \N^2} \{ J^2, T^a \} \ .
\end{eqnarray}
The corresponding $SU(3)$ analysis gives the Lagrangian
\begin{eqnarray}
\L^{\M} &&=
\sigma_B\, \Tr \left( \M \left( \Sigma + \Sigma^\dagger \right) \right)
\Tr \left(\overline B B \right)  
-\sigma_T
\, \Tr \left( \M \left( \Sigma + \Sigma^\dagger \right) \right)
\overline T^\mu T_\mu \nonumber\\  
&& +b_D\, \Tr \overline B \left\{ \left( \xi^\dagger \M \xi^\dagger
+ \xi \M \xi \right), B \right\} 
+b_F\, \Tr \overline B \left[ \left( \xi^\dagger \M \xi^\dagger 
+ \xi \M \xi \right), B \right] \nonumber\\
&& +c\, 
\overline T^\mu \left( \xi^\dagger \M \xi^\dagger + \xi \M \xi \right) T_\mu ,
\end{eqnarray}
in terms of two flavor-singlet parameters and three flavor-octet parameters.
%The $SU(3)$ parameters are related to the $1/N_c$ coefficients through
%\begin{equation}
%\sigma_B = {1 \over \sqrt{6}} \left( 3 c_0 
%+ {1 \over 4} c_2 \right), \quad
%\sigma_T = {1 \over \sqrt{6}} \left( 3 c_0 
%+ {5 \over 4} c_2 \right) 
%\end{equation}
%and
%\begin{eqnarray}\label{bdbfc}
%&&b_D = {1 \over 4} d_2, \nonumber\\
%&&b_F = {1 \over 2}  d_1 + {1 \over 6} d_2
%+{1 \over {12}} d_3,\\
%&&c = - {3 \over 2} d_1 - {5 \over 4} d_2
%- {5 \over 4} d_3 . \nonumber
%\end{eqnarray}

The leading order $1/N_c$ truncations for the flavor-singlet and flavor-octet
baryon amplitudes yield
\begin{eqnarray}
\H^0 = c_0 \N \openone \quad &\Rightarrow& \quad \sigma_T
= \sigma_B \\
\H^a = d_1 T^a 
\quad &\Rightarrow& \quad b_F = -\frac 1 3 c, \ b_D =0 \ , \nonumber
\end{eqnarray}
whereas the $1/N_c$ truncation of the flavor-octet baryon amplitude at first
subleading order in $1/N_c$ yields 
\begin{equation}
\H^a = d_1 T^a + d_2 {1 \over \N} \{ J^i, G^{ia} \}
\quad \Rightarrow \quad \left(b_D + b_F \right) = - \frac 1 3 c \ .
\end{equation}

In addition,
$U(3)$ planar flavor symmetry implies that the flavor-singlet amplitude 
${\cal H}^0$ is the ninth component of ${\cal H}^a$,
\begin{equation}
\H^0 = \frac 1 {\sqrt{6}} d_1 \N \openone
+ \frac 2 {\sqrt{6}}\left( d_2 
+ d_3 \right){1 \over \N} J^2 \ , 
\end{equation}
up to relative order $1/N_c$.  The corresponding nonet symmetry
relations for the $SU(3)$ parameters are
\begin{equation}
\sigma_B \rightarrow \frac 1 3 \left( 3 b_F -  b_D \right),
\ \sigma_T \rightarrow - \frac 1 3 c \ .
\end{equation}

\section{Calculating Chiral Loop Corrections}

The baryon propagator depends on the baryon spin through the hyperfine mass
operator.  It can be defined
in terms of the spin-$\frac 1 2$ and -$\frac 3 2$ 
projection operators
\begin{equation}
P_{\frac 1 2} = -\frac 1 3 \left( J^2 - \frac {15} 4 \right), \qquad
P_{\frac 3 2} = \frac 1 3 \left( J^2 - \frac {3} 4 \right) 
\end{equation}
%which satisfy
%\begin{equation}
%P_j P_j = P_j, \qquad P_{j^\prime} P_j =0 \ {\rm if} \ j^\prime \ne j \ .
%\end{equation}
as
\begin{equation}
{{i P_j} \over {(k^0 - \Delta_j)}}, \quad
\Delta_j = {1 \over {N_c}} \left( j (j+1) - j^\prime (j^\prime +1 ) \right)
\end{equation}
where the mass splitting $\Delta_j$ is given for
a propagating spin-$j$ baryon emitted with pions from a spin-$j^\prime$ baryon.
 
Chiral corrections can be calculated directly as products of $1/N_c$ baryon
operators.  The advantages of the $1/N_c$ operator expansion are that the group
theory and the $1/N_c$ cancellations are explicit.  A couple sample calculations
will be considered to illustrate the method.  

A general chiral loop calculation
involves a nonanalytic function $F(m, \Delta)$ of the meson masses $m$ and the
baryon hyperfine mass splitting $\Delta$.
First consider the leading order contribution to the flavor-${\bf 27}$
baryon mass splittings obtained from the 
baryon wavefunction renormalization diagram.
%For degenerate octet and decuplet baryons, the nonanalytic function reduces to
%$F(m, \Delta=0) \sim m^3/f^2$.  
The nonanalytic function can be decomposed into
flavor-singlet, octet and ${\bf 27}$ components.  The flavor-$\bf 27$ component is
given by
\begin{equation}
\Pi_{27}^{ab}= \left( \frac 1 3 F(\pi) - \frac 4 3 F(K) + F(\eta) \right)
\left( \delta^{a8} \delta^{b8} - \frac 1 8 \delta^{ab} 
- \frac 3 5 d^{ab8} d^{888} \right) \ .
\end{equation}
The wavefunction diagram yields a flavor-$\bf 27$ correction given by
\begin{equation} 
{1 \over {N_c}} \sum_{j= \frac 1 2 , \frac 3 2} 
\left( A^{ia} P_j A^{ib} \right) \Pi_{27}^{ab}(\Delta_j) \ .
\end{equation}
The flavor-$\bf 27$ Gell-Mann--Okubo mass combination,
$\frac 3 4 \Lambda + \frac 1 4 \Sigma - \frac 1 2 \left( N + \Xi \right)$,
is given by projecting this
chiral loop contribution onto the spin-$\frac 1 2$ baryons
\begin{equation}
{1 \over {N_c}} \left[P_{\frac 1 2} A^{i8} P_{\frac 1 2} A^{i8} P_{\frac 1 2}
F_{27}(m,0)
+P_{\frac 1 2} A^{i8} P_{\frac 3 2} A^{i8} P_{\frac 1 2} F_{27}(m,\Delta)\right]
\end{equation}
whereas the flavor-$\bf 27$ spin-$\frac 3 2$ baryon mass splitting
$-\frac 4 7 \Delta + \frac 5 7 \Sigma^* + \frac 2 7 \Xi^* - \frac 3 7 \Omega $, 
is given by the projection onto spin-$\frac 3 2$ baryons
\begin{equation}
{1 \over {N_c}} \left[P_{\frac 3 2} A^{i8} P_{\frac 1 2} A^{i8} P_{\frac 3 2}
F_{27}(m,-\Delta)
+P_{\frac 3 2} A^{i8} P_{\frac 3 2} A^{i8} P_{\frac 3 2} F_{27}(m,0)\right]\ .
\end{equation}
These complicated expressions reduce to linear combinations of the flavor-$\bf 27$
$1/N_c$ mass operators
\begin{equation}
{1 \over {N_c}} \left\{T^8, T^8 \right\} + {1 \over {N_c^2}}
\left\{T^8, \left\{J^i, G^{i8} \right\} \right\}\ .
\end{equation}

Next, consider the one-loop chiral correction 
$\delta A^{ia}$ to the baryon axial vector
flavor-octet current~\cite{fhjm} given by
\begin{equation}
\frac 1 2 F^{(1)}(m_b,0, \mu ) 
\left[ A^{jb}, \left[ A^{jb}, A^{ia} \right] \right] 
-\frac 1 2 F^{(2)}(m_b,0, \mu ) \left\{A^{jb}, \left[A^{ia}, \left[M, A^{jb}
\right] \right] \right\}
\end{equation}
up to corrections with more insertions of the mass operator $M$, where
\begin{equation} 
F^{(n)}(m_b, \Delta, \mu ) \equiv {{\partial^n F(m_b, \Delta, \mu )} \over
{\partial \Delta^n} } \ .
\end{equation}
The large-$N_c$ cancellations for the axial vector
currents occur only in the terms with zero or
one power of $M$.  Thus, a hybrid approach is suggested
in which the nonanalytic function is
expanded in a Taylor series about $\Delta =0$,
\begin{equation}  
F(m_b, \Delta) = F(m_b, 0) + F^{(1)}(m_b, 0) \Delta
+ \frac 1 2 F^{(2)}(m_b, 0) \Delta^2 + \tilde F(m_b, \Delta),
\end{equation}
and the leading terms are kept as $1/N_c$ operators while the remainder 
$\tilde F(m, \Delta)$ can be treated as a nonanalytic function in standard
HB$\chi$PT since it contains no large-$N_c$ cancellations.

\section{Conclusions}
A $1/N_c$ chiral Lagrangian for baryons has been formulated based on
large-$N_c$ QCD symmetries: 
$SU(3)_L \times SU(3)_R \times U(1)_V \times U(1)_A$ 
chiral symmetry
and contracted $SU(6)$ baryon spin-flavor symmetry.  Baryon chiral perturbation
theory using the $1/N_c$ chiral Lagrangian has a well-defined large-$N_c$ limit
and the chiral loop expansion satisfies the expected $1/N_c$ power counting for
quantum loops because of large-$N_c$ cancellations.  
Moreover, the spin, flavor and $1/N_c$
structure of the baryon chiral expansion is explicit in the $1/N_c$ 
operator expansion.

\section*{Acknowledgments}
This work was supported in part by the U.S. Department of Energy, under Grant
No. DOE-FG03-97ER40546, and by the Institute for Nuclear Theory at the
University of Washington.

\end{document}